\documentclass[prx,twocolumn,preprintnumbers,amsmath,amssymb,floatfix]{revtex4-1}
\usepackage{graphicx}
\usepackage{epstopdf}
\usepackage{dcolumn}
\usepackage{bm}
\usepackage{color}
\usepackage{natbib}
\usepackage{amsmath}
\usepackage{amssymb}
\usepackage{multirow} 

\usepackage{hyperref}

\hypersetup{colorlinks=true, linkcolor=blue, citecolor=red, urlcolor=magenta}

\begin{document}

\title
{Linear magnetoelectricity at room temperature in perovskite superlattices by design}

\author{Saurabh Ghosh} 
\author{Hena Das} 
\author{Craig J. Fennie } 
\email{fennie@cornell.edu} 
\affiliation{School of Applied and Engineering Physics, Cornell University, Ithaca, New York 14853, USA }

\date{\today}

\begin{abstract}
Discovering materials that display a linear magnetoelectric effect at room temperature is challenge. Such materials could facilitate novel devices based on the electric-field control of magnetism. Here we  present simple, chemically intuitive design rules to identify  a new class of bulk magnetoelectric materials based on the `bicolor' layering of $Pnma$ ferrite perovskites, e.g., LaFeO$_3$/ LnFeO$_3$ superlattices for which Ln  = lanthanide cation. We use first-principles density-functional theory calculations to confirm these ideas.  Additionally, we elucidate the origin of this effect and show it is a general consequence of the layering of any bicolor, $Pnma$ perovskite superlattice in which the number of constituent layers are odd (leading to a form of hybrid improper ferroelectricity) and Goodenough- Kanamori rules. Here, the polar distortions induce both weak ferromagnetism and a linear magnetoelectric effect. Our calculations suggest that the effect is 2-3 times greater in magnitude than that observed for the prototypical magnetoelectric material, Cr$_2$O$_3$. We use a simple mean field model to show that the considered materials order magnetically above room temperature. 
\end{abstract}
\maketitle

\section{Introduction}

Multiferroics are materials in which ferroelectricity and magnetism coexist.~\cite{Aizu1970, scottrev, spldnfb} Despite recent intense efforts to discover new multiferroics, there are surprisingly few materials that display this property at room temperature. Furthermore, the primary challenge remains to identify materials that have a functional coupling between an electrical polarization and a magnetization at room temperature
~\cite{Eom2008,Scott2013}. Such materials may, for example, facilitate technologically important devices based on the electric field control of magnetism~\cite{bibes,Maekawa2014,Dominique2007,Parkin1999,Albert2007,Ramesh2011}.

One way to design such cross-couplings is to start with a paraelectric material that is magnetically ordered and induce a ferroelectric lattice distortion.~\cite{Birol2012} For example, it was shown~\cite{Fennie2008} how a polar distortion -- in an antiferromagnetic--paraelectric (AFM-PE) material displaying linear magnetoelectricity -- would induce weak-ferromagnetism in the LiNbO$_3$ structure, e.g., FeTiO$_3$~\cite{Varga2012} or MnSnO$_3$~\cite{Angel2013}, and subsequently allow for the electric-field switching of the magnetization by 180$^{\circ}$.
Alternatively a ferroelectric distortion in an AFM-PE material that displays weak-ferromagnetism  can induce linear magnetoelectricity~\cite{Birol2012, Scott2013}.  Here, Bousquet and Spaldin recently realized that  the orthorhombic perovskites, space group \textit{Pnma}, are prime realizations and proposed epitaxial strain as a route  to induce ferroelectricity~\cite{bousquet11}. They showed from first-principles that under large strain, \textit{Pnma} CaMnO$_3$ indeeds becomes ferroelectric. The  polar lattice distortions lowers the symmetry to 
$Pmc2_1$ and a linear magnetoelectric effect is subsequently induced.

In the present study, we  take an alternate route to achieve ferroelectrically induced  linear magnetoelectricity in \textit{Pbnm} (space group number 62, in standard setting which is $Pnma$) perovskites by taking advantage of a recent direction~\cite{bousquet08,benedek11}  whereby  the combination of  rotations/tilts of the BO$_6$ octahedra and A-site cation ordering facilitate  ferroelectric order~\cite{rondinelli12, benedek12, mulder13}, without the need for strain.
We consider the rare-earth (La/Ln)Fe$_2$O$_6$ orthoferrite superlattices in which the La and Ln cations (Ln = Ce, Nd, Sm, Gd, Dy, Tm, Lu  and Y) are ordered in layers along the crystallographic \textit{c}-axis, where respective supercells have been constructed as $\sqrt{2}a_p ~\times~\sqrt{2}a_p~\times2a_p$ (here, $a_p$ the pseudocubic lattice parameter of $Pbnm$ LaFeO$_3$).
Note that similar results are obtained for (LaFeO$_3$)$_n$/(LnFeO$_3$)$_m$ heterostructures of \textit{Pnma} materials when both\textit{ n} and \textit{m} are odd~\cite{Rosseinsky2014,AtomisticHIF}. The choice of the orthorferrites was dictated by the fact that bulk  LnFeO$_3$ materials order magnetically  above room temperature, with T$_N$ as high as $\sim$ 740 K for LaFeO$_3$~\cite{White69,pnmamag1,marezio70,seo08}.

We show from first-principles that the magnitude of the linear magnetoelectric (ME) tensor in these  heterostructures is 2-3 times that of the canonical linear ME, Cr$_2$O$_3$.~\cite{ini2008,ini2009,vanderbilt2012,Iyama2013,Kris2009} This work provides a practical route to create a new class of multiferroic materials that display a linear magnetoelectric effect at room temperature whereby  octahedral rotations mediate a nontrivial coupling between magnetism and ferroelectricity~\cite{benedek11,benedek12,lawes11,ghosez11,zanolli13}.

\section{Computational Details}

First-principles calculations have been carried out using density functional theory~\cite{DFT} with projector augmented wave (PAW) potentials~\cite{paw} and within LSDA+U~\cite{anisimov97}, as implemented in the Vienna ab initio simulation package (VASP)~\cite{vasp}. We have considered PAW potentials for Ln$^{3+}$ ions where $f$-states are treated in the core, eliminating magnetic orderings associated with the $f$-state magnetism which occurs at much lower temperature. For Fe$^{3+}$ ions, we have included the on-site $d-d$ Coulomb interaction parameter U=6.0 eV, and exchange interaction parameter J=1.0 eV. The exchange-correlation part is approximated by PBEsol functional~\cite{pbesol}, which improves the structural descriptions over standard LDA or GGA~\cite{wahl08}. The convergence in total energy and Hellman-Feynman force were set as 0.1 $\mu$eV and 0.1meV$/$\AA, respectively. All calculations have been performed with a 500 eV energy cutoff and with a $\Gamma$-centered $6\times6\times4$ k- point mesh. convergence has been tested with higher energy cutoff, k-mesh and found to be in agreement with the present settings. Non-collinear magnetization calculations were performed with L-S coupling~\cite{lscoupling}, whereas total polarization was calculated with the Berry phase method~\cite{kingsmith94} as implemented in VASP.

\section{Polarization, Magnetization and Switching}
\begin{figure}[h]
\centering\vspace{1pt}
\includegraphics[scale=0.85]{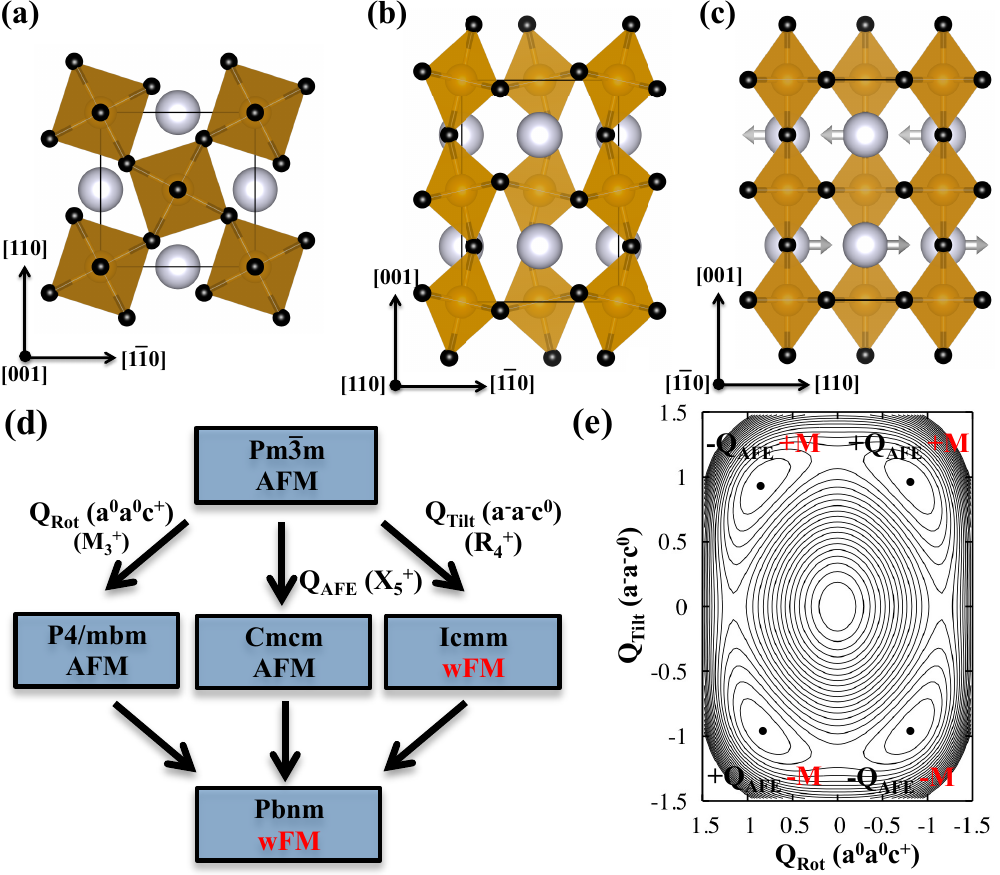}  
\caption {The `orthorhombic' ABO$_3$ perovskite, space group $Pbnm$. The structure, Glazer pattern $a^-a^-c^+$, is described by three symmetrize  basis modes of cubic \textit{Pm$\bar{3}$m:} (a) Q$_{Rot}$, in-phase rotation of BO$_6$ octahedra about [001] (irrep. M$_3^+$), (b) Q$_{Tilt}$, tilt of BO$_6$ octahedra about [110] with A-site displacement (irrep. R$_4^+$), (c) Q$_{AFE}$ anti-ferroelectric  A-site displacement (irrep. X$_5^+$). (d) The group-subgroup relation from high-symmetric  \textit{Pm$\bar{3}$m} to low-symmetric \textit{Pbnm}. (e) Two dimensional energy surface contour of LaFeO$_3$ with respect to the primary Q$_{Rot}$ ($a^0a^0c^+$) and Q$_{Tilt}$ ($a^-a^-c^0$) distortions.}
\label{Fig1}
\end{figure}

The space group symmetry of the orthorhombic $Pbnm$ structure adopted by most perovskites~\cite{King07,King10} is established by two symmetry-lowering structural distortions of the cubic \textit{Pm$\bar{3}$m} perovskite structure: an in-phase rotation of the BO$_6$ octahedra about the cubic [001] axis (transforming like the irreducible representation M$_3^+$) Q$_{Rot}$ and an out-of-phase tilt of the BO$_6$ octahedra about the cubic [110] axis (transforming like the irreducible representation R$_4^+$), Q$_{Tilt}$, as shown in Figure~\ref{Fig1}a and Figure~\ref{Fig1}b, respectively. Together these two distortions produce the Glazer rotation pattern $a^-a^-c^+$.
Other kinds of structural distortions are also allowed by symmetry in the  \textit{Pbnm} structure. In particular, recent work has shown that anti-polar displacements of the A-site cations $-$ as shown in  Figure~\ref {Fig1}c, the displacements are equal in magnitude but in opposite directions in adjacent AO planes -- play a crucial role in stabilizing the structures of $Pbnm$ perovskites~\cite{woodward97a,benedek13}.
 These anti-polar displacements (transforming like the irrep X$_5^+$) are coupled to the two rotation distortions, \textit{i.e.}, there is a tri-linear term in the free energies of cubic \textit{Pm$\bar{3}$m} perovskites that couples all three distortions~\cite{amisi12}, $\mathcal{F}$ = Q$_{\rm AFE}$ Q$_{\rm Tilt}$  Q$_{\rm Rot}$, where Q$_{\rm AFE}$, Q$_{\rm Tilt}$, and Q$_{\rm Rot}$ are the amplitudes of the anti-polar distortion, tilt and rotation distortions respectively.
 Hence, reversing the sense of either Q$_{\rm Tilt}$ or Q$_{\rm Rot}$  will therefore reverse the direction of the anti-polar displacements, Q$_{\rm AFE}$. 
 $Pbnm$ orthoferrites, the Fe spins typically order in a G-type antiferromagnetic ordering pattern with weak ferromagnetism (wFM) along the \textit{Pbnm} orthorhombic c-axis. This wFM is in fact induced by the Q$_{\rm Tilt}$ distortion in \textit{Pbnm} and hence the sense of this particular rotation and the direction of the canted magnetic moment are naturally coupled in a non-trivial way. These ideas are summarized in Figures~\ref{Fig1}d and e.

\begin{figure}
\centering
\includegraphics[scale=0.40]{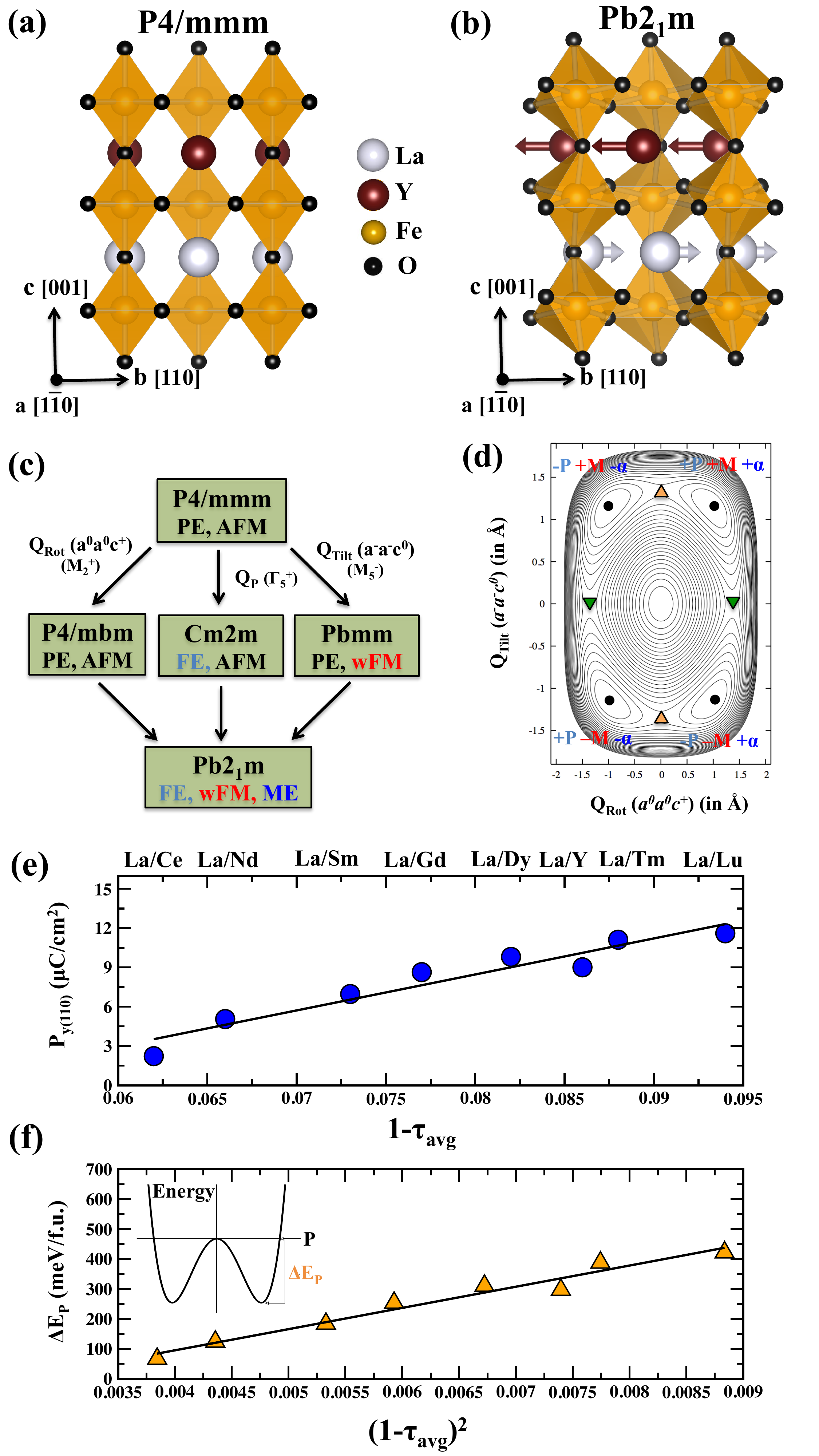} 
\caption {Structural and ferroelectric properties of (La/Ln)Fe$_2$O$_6$ superlattices. (a)  (La/Y)Fe$_2$O$_6$ superlattice in high-symmetry $P4/mmm$ structure,  (b) lowest energy $Pb2_{1}m$ structure with rotation and tilt of FeO$_6$ octahedra, (c) Group-subgroup relation from $P4/mmm$ to other lower energy structures, (d) Two dimensional energy surface contours for (La/Y)Fe$_2$O$_6$ superlattice with respect to the primary Q$_{Rot}$ ($a^0a^0c^+$, irrep. M$_2^+$) and Q$_{Tilt}$ ($a^-a^-c^0$, irrep. M$_5^-$) distortions. In contour, Black dots represents the the possible minimum structures, each corresponding to different sense of FeO$_6$ octahedra rotations, Green triangles (down) indicate to the $P4/mbm$ structures where Q$_{Tilt}$ is zero and Orange triangles (up) indicate to the $Pbmm$ structure where Q$_{Rot}$ is zero.  Variation of (e) Polarization P and  (f) Switching barriers along Rotation ($\Delta E_{Rot}$) with respect to (1-$\tau_{avg}$) and (1-$\tau_{avg}$)$^2$, respectively.}
\label{Fig2}
\end{figure}

The $Pbnm$  perovskites are thus a system in which octahedral rotations mediate a non-trivial coupling between antiferroelectricity and magnetism. The recently developed theory of hybrid improper ferroelectricity has shown how antiferroelectricity in $Pbnm$ perovskites can give rise to ferri-electricity in perovskite hetereostructures,~\cite{mulder13} such as (A/A')B$_2$O$_6$ double perovskites, Figure~\ref{Fig2}a. 
A simple picture~\cite{benedek12} that elucidates the mechanism is the following: the two rotation distortions, Q$_{\rm Rot}$ and Q$_{\rm Tilt}$, break inversion symmetry at the A-site of the cubic  \textit{Pm$\bar{3}$m} structure whereas the A/A' cation ordering breaks B-site inversion symmetry, such that the A-site displacements depicted in Figure~\ref{Fig1}c are no longer equal and opposite but instead give rise to a macroscopic polarization, as shown in  Figures~\ref{Fig2}b (in other words, this A-site displacement mode becomes a zone-center polar mode in the cation ordered unit cell, which has $P4/mmm$ symmetry in the absence of any rotations,  Figure~\ref{Fig2}a). The key is that since the (now polar) A-site displacements are coupled to the rotations as described above, switching the direction of the polarization, Q$_P$, will switch the sense of one of the rotations. If it is Q$_{\rm Tilt}$ that switches, then the direction of the canted moment will also switch, resulting in electric field control of the magnetization. 
These facts are summarized in  Figure~\ref{Fig2}c and d (it is highly instructive to compare  Figures \ref{Fig1}d  and \ref{Fig2}c.)

We have used first-principles total energy calculations to consider the complete manifold of possible lower symmetry structures for this class of compounds and have identified the structure shown in  Figure~\ref{Fig2}b as the lowest in energy. This structure has polar $Pb2_1m$ space group symmetry and displays both ferroelectricity (with a polarization along the orthorhombic y axis, P$_y$) and weak ferromagnetism (with a net magnetization along the z-axis, M$_z$). The resultant magnetic configuration has magnetic point group $m^{\prime}m2^{\prime}$ and consists of G-type AFM ordering with the easy axis along $x$, A-type AFM ordering along the $y$-axis and a FM canting of spins along the $z$-axis (G$_x$, F$_z$)~\cite{pnmamag1, marezio70,seo08,pnmamag2, pnmamag3, pnmamag4, pnmamag5}.

Since, the origin of the polarization in our  (La/Ln)Fe$_2$O$_6$ materials is  a non-cancelation of the LaO and LnO  layer polarizations. A simple way to increase this non-cancellation and hence the polarization is to choose a Ln cation whose tendency to off-center from  the ideal perovskite A-site differs greatly from that of La. This is accomplished by choosing a Ln cation that is much smaller than La. 
In Figure~\ref{Fig2}e, we  show that the magnitude of  P$_y$ monotonically increases as the Ln cation becomes smaller, from 2.2 $\mu$C/cm$^2$ for Ln=Ce to 11.6 $\mu$C/cm$^2$ for Ln=Lu.  Said another way, as the average tolerance factor $\tau_{\rm avg}$ decreases, the polarization increases (note, $\tau_{ABO_3}= (r_A+ r_O) / \sqrt2 (r_B + r_O )$, where $r_A$, $r_B$ and $r_O$, are ionic radii of A, B and O atoms respectively). As discussed by Mulder et al.~\cite{mulder13}, such a simple behavior is only true when one of the two A-site cations is the same for all compounds.  Additionally,  M$_z$ is  roughly 0.07 $\mu_B$/f.u$.$ for all compounds (there is a small increase in M$_z$ as $\tau_{\rm avg}$ decreases, 0.065-0.070$\mu_B$/f.u, but is too small to be significant).

The direction of the polarization can in principle  be switched 180$^{\circ}$ between symmetry equivalent states with the application of an electric-field. In this process the sense of either   Q$_{\rm Rot}$ or Q$_{\rm tilt}$ will switch. The question as to which distortion would actually switch is a challenging, dynamical problem, one for which today we still don't have a satisfactory answer (please see  Ref.~\cite{zanolli13} for a nice discussion) and beyond the scope of this paper. We know, however, that switching does depend in some way on the energy barriers between the energy minima displayed in  Figure~\ref{Fig2}d. Understanding how to control any of the energy barriers is useful information, even if the precise path is not known.  Within this limited sense let us briefly discuss the naive switching paths.

We found  that the barrier height along  Q$_{\rm Rot}$, $\Delta$E$_{\rm Rot}$= $E_{Pb2_1m}$ - E$_{Pbmm}$,  is about three times smaller than the  barrier height along Q$_{\rm Tilt}$, $\Delta$E$_{\rm Tilt}$= $E_{Pb2_1m}$ - E$_{P4/mbm}$ for all the compounds we considered. From this we conclude that it is more likely that  Q$_{\rm Rot}$ would switch when the polarization switches.  Since M$_z$ switches only if Q$_{\rm Tilt}$ switches, the (La/Ln)Fe$_2$O$_6$ systems do not appear to be likely candidates to pursue the electric-field switching of the magnetization. Furthermore, examination of Figure~\ref{Fig2}e shows that the ideal energy barrier to switch the polarization, $\Delta E_{P} \equiv \Delta E_{Rot}$,  increases dramatically as $\tau_{\rm avg}$ decreases, as expected from the design rules of Ref.~\cite{mulder13}. In fact, it is not likely that the polarization in the majority of these materials could ever be switched under realistic electric field strengths, other than perhaps (La/Ce)Fe$_2$O$_6$, which has the lowest switching barrier (63 meV/f.u$.$ along the  $a^0a^0c^+$ rotation path).

\section{Linear magnetoelectric coupling}

\begin{figure}[h]
\centering
\includegraphics[scale=0.43]{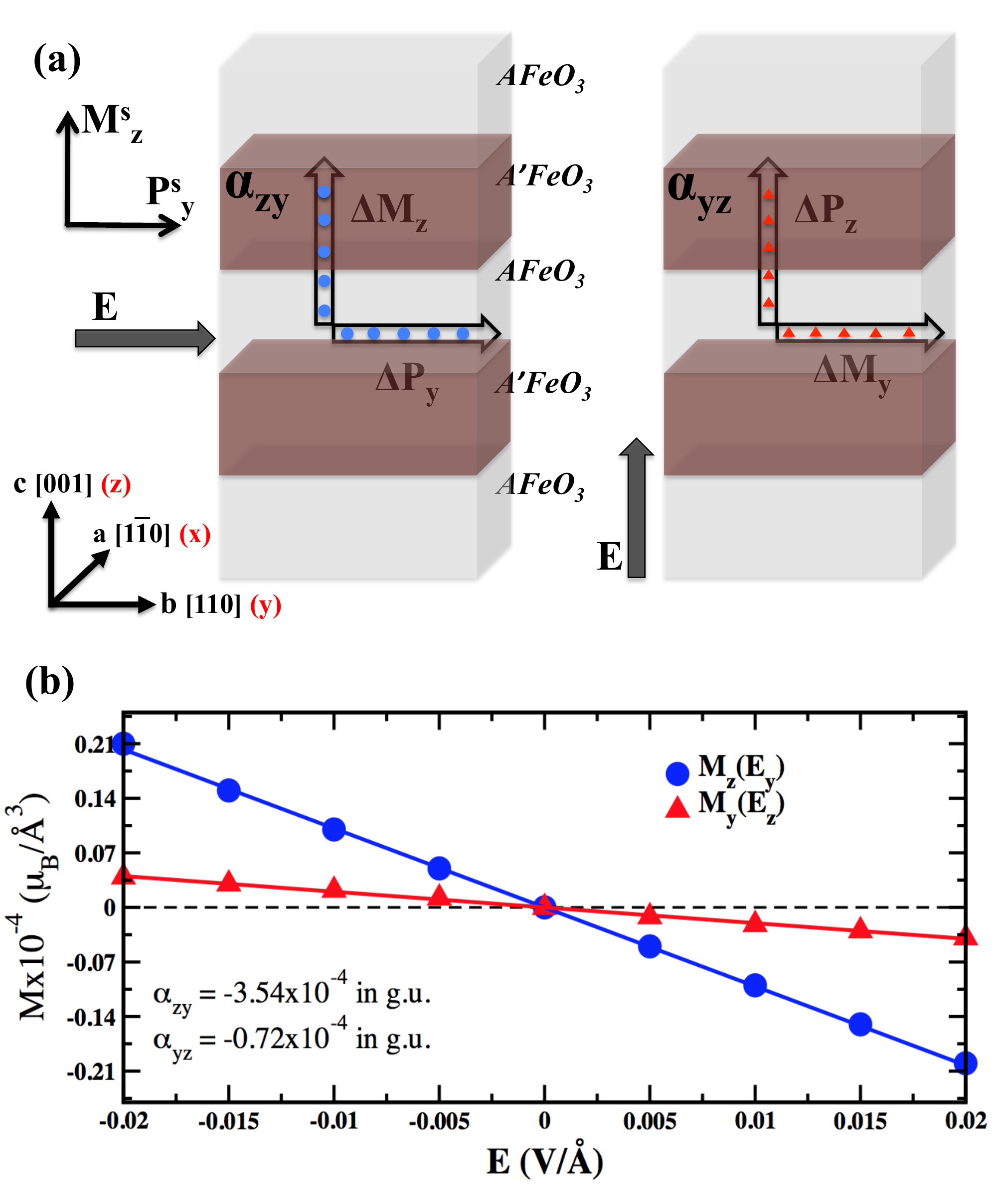} 
\caption {Linear Magnetoelectric response of (La/Y)Fe$_2$O$_6$ superlattice. (a) Superlattice with La/Y cation ordering along $z$-direction which is the crystallographic $c$-axis. The spontaneous HIF polarization is along \textit{y}-direction and net magnetization M in the system along $z$-direction(P$^s_y$, M$^s_z$). The ME tensor has two non-zero components. First, $\alpha_{yz}$, change in polarization along $z$-direction ($\Delta$P$_z $) changes the magnetization along $y$-direction ($\Delta$M$_y$) and second $\alpha_{zy}$, change in polarization along $y$-direction ($\Delta$P$_y$) changes the magnetization along $z$-direction ($\Delta$M$_z$), (b) Change in magnetization (with respect to the saturation magnetization) subject to the variation of electric field (\textbf{E}). Linear magnetoelectric components $\alpha_{zy}$ and $\alpha_{yz}$ are obtained form the slope of M$_z$ vs. E$_y$ and M$_y$ vs. E$_z$ curve, respectively. The magnitude of $\alpha_{zy}$ and $\alpha_{yz}$ are given in Gaussian units (g.u.).}
\label{ME}
\end{figure}

The structural distortions associated with the spontaneous polarization, however, induce by design a linear magnetoelectric effect (which does not require switching of either the polarization or the magnetization),
\begin{eqnarray}
\Delta {\rm M}_i &= & \Sigma \alpha_{ij} E_j   \\  
\Delta {\rm P}_i &= & \Sigma \alpha_{ij} H_j ,  
\label{magme}
\end{eqnarray}
where $\Delta$M$_i$ ($\Delta$P$_i$) is the induced  magnetization (polarization) along  the  \textit{i}th direction due to an  electric (magnetic) field applied along the  \textit{j}th direction.  The magnetic point group of all  (La/Ln)Fe$_2$O$_6$ compounds is m$^{\prime}$m2$^{\prime}$, therefore, the only non-zero components of the linear ME tensor are, 
\begin{equation}
\mathbf{\alpha} =\begin{bmatrix}
0&0&0\\
0&0&\alpha_{yz}\\
0&\alpha_{zy}&0\\
\end{bmatrix}
\label{alpha}
\end{equation}
where  $\alpha_{yz} \neq \alpha_{zy}$ (the ME process associated with these components are schematically shown in Figure~\ref{ME}(a)). 

The design strategy guarantees the existence of  $\alpha$, but what is its magnitude? Here, we used the method described in Ref.~\cite{ini2008} to calculate the lattice contribution of $\alpha$ (although the  linear ME response can have both lattice and electronic contributions,~\cite{Birol2012, ini2008} the method of Ref.~\cite{ini2008} should give a reasonable order of magnitude estimate).

A brief description of this method is the following:  considering only the lattice contribution to the energy, the energy of the \textit{Pbnm} crystal ($\mathcal{U}$) under an applied electric field ($\textbf{E}$) can be given by,
\begin{equation}
\mathcal{U} (\textbf{q}_n, \textbf{E}) = \mathcal{U}_0 + \frac{1}{2}  \displaystyle\sum_{n} C_n q{_n^2} - \displaystyle\sum_{n} q_n  \textbf{p}{_n}\cdot\textbf{E}
\label{energy}
\end{equation}
where,  $q_n$, C$_n$, and \textbf{p}$_{n}$  are the amplitude,  force constant, and dielectric polarity of the $n^{th}$  infrared (IR) active force constant eigenvector, $\hat{\textbf{q}}_n$, respectively. The dielectric polarity \textbf{p}$_{n}$ of the $n^{th}$-IR active mode can be calculated as, $\textbf{p}{_n}= {\partial P_n}/{\partial q_n}$, where P is the polarization (note that the force constants and dielectric polarity are routinely calculated from first principles.~\cite{ini2008,Das2013})
Therefore, for a given electric field it is straightforward to calculate the induced atomic displacements associated with each force constant eigenvector, $\textbf{q}_n= q_n \hat{\textbf{q}}_n$, where  $q_n= \frac{1}{C_n}\textbf{p}{_n}\cdot \textbf{E}$. 
Subsequently, the linear ME tensor, 
\begin{equation}
\alpha_{ij} = {\partial M_i}/{\partial E_j},  
\end{equation}
can be calculated by freezing in the total induced atomic displacements, $\mathbf{u}= \sum_{n} q_n \hat{\textbf{q}}_n$, then recalculating the net magnetization.

As an example let us discuss the calculation of the  linear ME response for (La/Y)Fe$_2$O$_6$  (P$^s_y$=9.0 $\mu$C/cm$^2$ and P$_x$=P$_z$=0; M$^s_z$ = 0.13 $\mu_B$ and M$_x$=M$_y$=0).  It is useful to keep in mind that in $Pb2_1m$  the IR modes transform as irreducible representations $\Gamma_1$, $\Gamma_3$, or $\Gamma_4$, each leading to a polarization along the $y$, $z$, and $x$ directions, respectively. By symmetry, the $\Gamma_4$ modes do not mediate a ME effect. This corresponds to the fact that  $\alpha_{ij}$ for any $i$ or $j$ = $x$ is zero by symmetry. For the purpose of the calculation, we imagine the experiment in which an electric-field is applied and the resulting change in magnetization is measured.

With the application of an electric field along the $y$ direction, symmetry dictates that only the $\Gamma_1$  modes respond, i.e.,  $\textbf{p}{_n}\cdot\textbf{E} =  \textbf{p}{_n}\cdot\hat{\textbf{y}}E  \ne 0$ for  $n\in \Gamma_1$.  The induced atomic displacements 
\begin{equation}
\textbf{u}_{\Gamma_1} = \displaystyle\sum_{n\in \Gamma_1} \dfrac{1}{C_n}p_nE_y \hat{\textbf{q}}_n,
\end{equation}
were frozen into the equilibrium structure and the change in  magnetization, which by symmetry is  along the $z$ direction, was calculated from first-principles. This procedure was repeated for various magnitudes of the applied electric field. These results are shown in Figure~\ref{ME}(b)),   the slope of which gives the linear ME coupling $\alpha_{zy}$. We find that the magnitude of $\alpha_{zy}$ is 3.54x10$^{-4}$ g.u., which is 2-3 times larger than the transverse linear ME response of the prototype ME compound Cr$_2$O$_3$~\cite{ini2008} at 0 K.

With the application of an electric field along the $z$ direction, symmetry dictates that only the  $\Gamma_3$ modes respond, i.e., $\textbf{p}{_n}\cdot\textbf{E} =  \textbf{p}{_n}\cdot\hat{\textbf{z}}E  \ne 0$ for  $n\in \Gamma_3$.  The induced atomic displacements 
\begin{equation}
\textbf{u}_{\Gamma_3} = \displaystyle\sum_{n\in \Gamma_3} \dfrac{1}{C_n}p_nE_z \hat{\textbf{q}}_n.
\end{equation}
can be calculated. Here we find that the corresponding linear ME response (see Figure~\ref{ME}(b)), measured by the component $\alpha_{yz}$ is equal to 0.72x$10^{-4}$ g.u., much weaker than $\alpha_{zy}$.

\section{Electronic Structure} 

\begin{figure}[h]
\begin{center}
\includegraphics[scale=0.41]{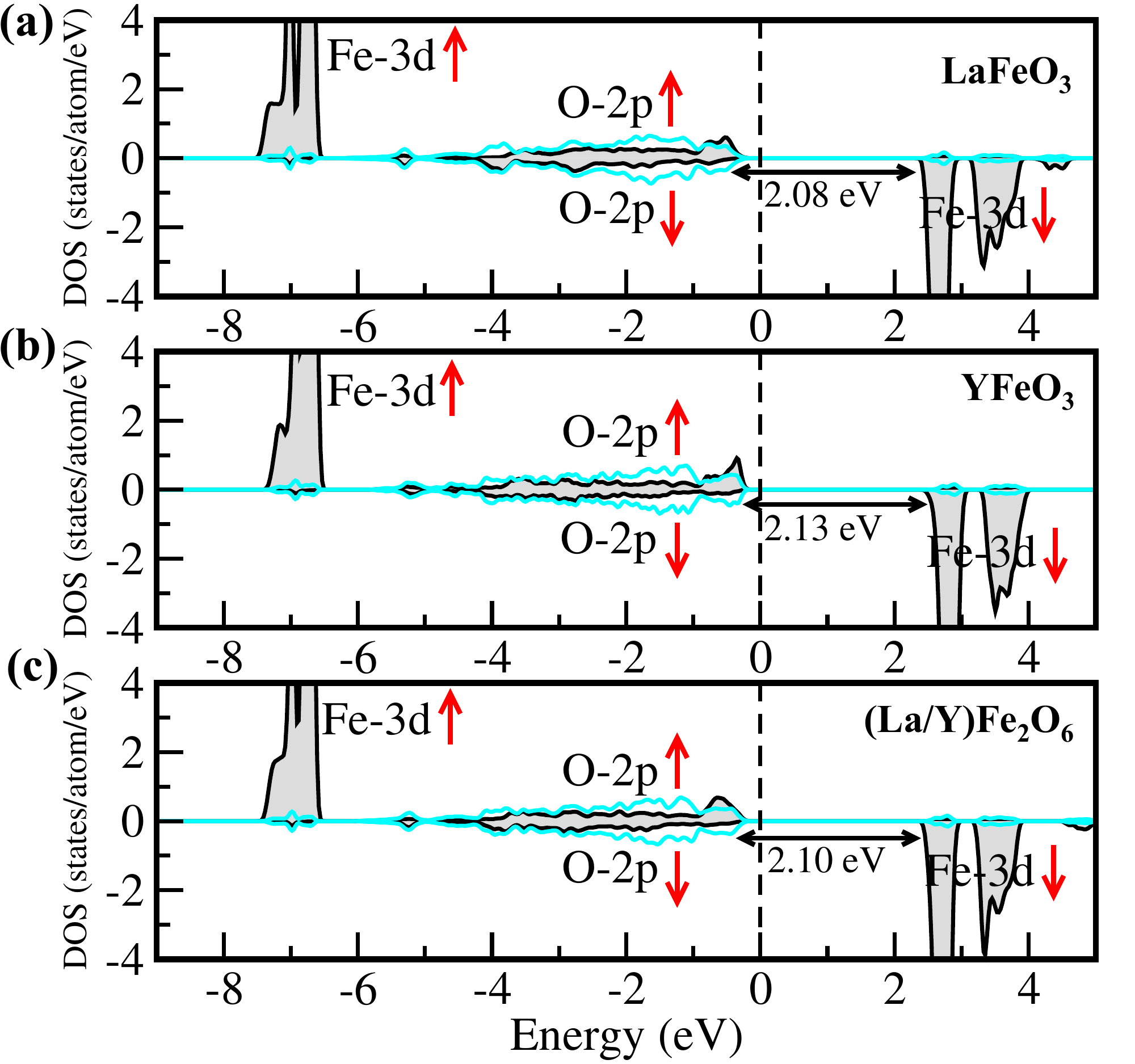}  
\end{center}
\caption{Calculated Density of States (DOS) for (a) LaFeO$_3$, (b) YFeO$_3$ in $Pbnm$ and (c) (La/Y)Fe$_2$O$_6$ in $Pb2_1m$ symmetry. For Fe$^{+3}$ ions, we have used the on-site $d-d$ Coulomb interaction parameter U=6.0 eV and exchange interaction parameter J=1.0 eV.}
\label{DOS}
\end{figure}

We have been discussing an approach to create room temperature linear magnetoelectrics by layering \textit{Pbnm} materials. From previous work, the robustness and universality of the polar ground state is clear. There are, however, two important questions concerning the magnetic state that need to be addressed in order to support the claim of room temperature magnetoelectricity. 

The first concerns the type of antiferromagnetic ordering. As we discussed, we have found by direct first-principles calculations that all of the superlattices we considered order in a G-type pattern, which is required to observe the ME physics.  There is a simple reason why this should be the case as the G-type antiferromagnetic  ground state of the orthoferrites is driven by the electronic configuration of the Fe$^{+3}$ ion.
As an example, let us briefly discuss the basic electronic structure of $Pbnm$ LaFeO$_3$ and YFeO$_3$ and the superlattice made out of them \textit{i.e}. (La/Y)Fe$_2$O$_6$ in 
$Pb2_1m$ symmetry. As shown in Figure~\ref{DOS}(a) and (b), both LaFeO$_3$  and YFeO$_3$  are charge transfer insulators. The valence band is formed by majority Fe-3$d$ states and O-2$p$ states, while the minority Fe-3$d$ states are completely empty and form the conduction band. Due to the $d^5$ electronic configuration only antiferromagnetic superexchange interactions between Fe$^{+3}$ ions via single O-$2p$ orbitals are allowed. The ferromagnetic contribution involving two perpendicular $2p$ orbitals is negligibly small as the Fe-O-Fe bond angle is close to 180$^{\circ}$. Therefore, the G-type AFM configuration is universal for LnFeO$_3$ systems.
 We have found that the layered arrangement of La/Y cations results in negligible changes to the basic electronic structure of the energy level positioning and band width (see Figure~\ref{DOS}(c)). We therefore expect that the major component of Fe spins in the  ground state of the (La/Y)Fe$_2$O$_6$  superlattice will be  G-type (it is important to note that  the G-type AFM configuration in $Pb2_1m$, with the magnetic anisotropy along any crystallographic axis, allows linear ME coupling).

\section{ Ordering Temperature} 

\begin{figure}[h]
\begin{center}
\includegraphics[scale=1.0]{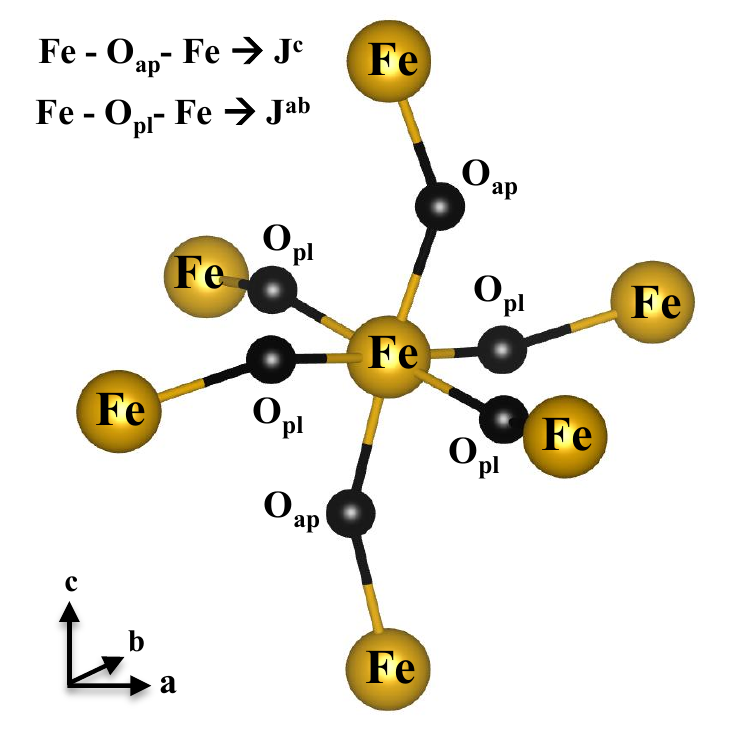}  
\end{center}
\caption{Exchange Interactions pathways, $J_{nn}^{ab}$ implies nearest neighbor Fe-Fe interactions (Fe- Fe distance 3.85 \AA~in $ab$ plane) mediated by planer Oxygen (O$_{pl}$), whereas  $J_{nn}^{c}$ implies nearest neighbor Fe-Fe interactions (Fe-Fe distance 3.76 \AA~along $c$-axis) mediated by apical Oxygen (O$_{ap}$). }
\label{J}
\end{figure}

The second important question that needs to be addressed is whether or not we expect the spins to order at room temperature. A relatively straightforward mean field approach to calculating the N\'{e}el temperature (T$_N$) involves mapping total energy calculations onto a Heisenberg model, from which the magnetic exchange interactions, $J_{ij}$, can be extracted.
Unfortunately, it is well know that the results obtained through this approach depend sensitively on the particular value of Hubbard U. Here we can take advantage of the fact that the experimental values of T$_N$ are known for the perovskite constituents of our superlattices.
Table1, shows the calculated values of T$_N$ for LaFeO$_3$, YFeO$_3$, and the (La/Y)Fe$_2$O$_6$ superlattice for a fixed value of U, considering up to the third nearest neighbor exchange interactions. There are a few things to note. 
 We found that  the dominant interaction is only between nearest neighbor ($nn$) spins (the interaction pathways are shown in Figure ~\ref{J}) and the average $nn$ exchange interaction ($J_{nn}^{avg}$) of (La/Y)Fe$_2$O$_6$ is almost equal in value to that of LaFeO$_3$. Additionally, when compared to the experimental values,  our calculations generally overestimated T$_N$, however, the calculated  ratio of $T_N^{LaFeO_3}$ and $T_N^{YFeO_3}$ is in good agreement with the ratio of the measured values.  Given this fact,  the $T_N$ of (La/Y)Fe$_2$O$_6$ superlattice is expected to be close to the magnetic transition temperature of LaFeO$_3$ ($\sim$ 740K). For the other (La/Ln) superlattices, we have also calculated the corresponding $T_N$ and found that for all the cases it is around the T$_N$ of LaFeO$_3$. This leads us to propose that these superlattices are expected to order magnetically above room temperature.

\begin{table}
\begin{center}
\caption{Computed superexchange constants ($J_{ij}$) and mean field estimated magnetic transition temperature. $J_{nn}^{ab}$ and $J_{nn}^{c}$ represent nearest neighbor Fe-Fe interaction mediated via planner and apical oxygen, respectively. The average interaction is given by $J_{nn}^{avg}=(4\times J_{nn}^{ab}+2\times J_{nn}^c$)/6. See Figure~\ref{J}).}
\begin{tabular}{cccc|c|c}
\hline
\hline
System&$J_{nn}^{ab}$&$J_{nn}^c$&$J_{nn}^{avg}$&\multicolumn{2}{c}{T$_N$ (K)}\\
&&&&Computed&Experiment\\
\hline
LaFeO$_3$&5.81&5.20&5.61&1139&740~\cite{pnmamag1}\\
YFeO$_3$&5.20&4.51&4.97&1009&655~\cite{pnmamag1}\\
(La/Y)Fe$_2$O$_6$&5.60&5.30&5.50&1117&--\\
\hline
\hline
\end{tabular}
\end{center}
\label{TJ}
\end{table}
\section{Conclusion} 
We have used first-principles calculations to identify a family of (La/Ln)Fe$_2$O$_6$ superlattices that may display a strong linear magnetoelectric effect at room temperature. Although the magnetoelectricity is ferroelectrically induced, polarization switching is not required to observe the effects studied here. An advantage of the superlattice approach is the possibility for additional functionality over that of the strain-induced approach due to the natural, non-trivial coupling of the polar, magnetic and rotation domains within the hybrid improper mechanism. Although the (1/1) materials will be challenging to synthesize, even as thin-films, similar results should be observed for any (LaFeO$_3$)$_n$/(LnFeO$_3$)$_m$ superlattice when both\textit{ n} and \textit{m} are odd~\cite{benedek12, Rosseinsky2014}.

\section{acknowledgement}
The authors acknowledge useful discussions with Nicole Benedek, James Rondinelli, Philippe Ghosez, and Darrell Schlom. S.G. and C.J.F$.$ we supported by the Army Research office under grant No$.$ W911NF-10-1-0345, and H.D$.$ was supported by the NSF-MRSEC Center for Nanoscale Science at Penn State, DMR-0820404.

\end{document}